\documentclass[%nofootinbib
 reprint,
superscriptaddress,
 amsmath,amssymb,
 aps,
 prl,
floatfix]{revtex4-2}

\usepackage{graphicx}% Include figure files
\usepackage{float} 
\graphicspath{{images/}}
\usepackage{dcolumn}% Align table columns on decimal point
\usepackage{bm}% bold math
\usepackage{amssymb}
\usepackage{soul}
\usepackage[dvipsnames]{xcolor}
\usepackage[normalem]{ulem}

\begin{document}

\preprint{APS/123-QED}
%%TC:ignore
\title{Enhancement of mechanical squeezing via feedback control}
% Force line breaks with \\

\author{Chao Meng}
\email{chao.meng@nbi.ku.dk}
\affiliation{%
Niels Bohr Institute, University of Copenhagen, 2100 Copenhagen, Denmark}%

\author{Warwick P. Bowen}
\email{wbowen@physics.uq.edu.au}
\affiliation{%
 Australian Research Council Centre of Excellence in Quantum Biotechnology, School of Mathematics and Physics, University of Queensland, St Lucia, Queensland 4072, Australia
}%

\date{\today}% It is always \today, today,

\begin{abstract}

We explore the generation of nonclassical mechanical states by combining continuous position measurement and feedback control. We find that feedback-induced spring softening can greatly enhance position squeezing. Conversely, even with a pure position measurement, we find that spring hardening can enable momentum squeezing.  Beyond enhanced squeezing, we show that feedback also mitigates degradation introduced by background mechanical modes. Together, this significantly lowers the barrier to measurement-based preparation of nonclassical mechanical states at room temperature.

\end{abstract}
%%TC:endignore
\maketitle
% \tableofcontents

Measurements play a crucial role in the preparation of quantum states and are utilized across a variety of fields, including quantum computing~\cite{riste_deterministic_2013}, quantum sensing~\cite{cox_deterministic_2016,sayrin_real-time_2011}, and tests of fundamental physics~~\cite{minev_catch_2019}. Continuous measurement of the position of a mechanical oscillator has been extensively studied and has established a standard limit to measurement precision~\cite{braginsky_quantum_1980}, impacting precision optomechanical sensors and gravitational wave detectors~\cite{muller-ebhardt_quantum-state_2009,miao_achieving_2010}. 

Recently, it has been shown that a position squeezed mechanical state can be prepared by continuous measurement as long as the measurement rate is faster than the rate that noise couples to the position of the oscillator~\cite{meng_mechanical_2020}. The noise coupling rate is related to the mechanical frequency. At room temperature, this necessitates the use of a low frequency mechanical resonance. Typical high-frequency mechanical oscillators (above a megahertz) used in optomechanical experiments~\cite{de_lepinay_quantum_2021,kotler_direct_2021,rossi_measurement-based_2018,guo_feedback_2019,aspelmeyer_cavity_2014,wilson_measurement-based_2015} cannot enter the squeezing regime even with state-of-the-art technology. Moreover, background of mechanical resonances can severely degrade state preparation~\cite{meng_measurement-based_2022}. This degradation cannot be evaded by increasing the measurement rate as this also increases the strength of background interactions. Together, this raises the question of whether it is practical to generate quantum squeezing at room temperature via continuous measurement. Furthermore, although momentum squeezing is crucial for enhancing force sensing, it is not clear how to achieve this via position measurement.

In this Letter, we show that the above-mentioned limitations can be overcome through feedback control. We find that feedback-induced spring softening, shifting the mechanical resonance frequency lower, can improve conditional position squeezing by a factor equal to the fractional change in resonance frequency. Moreover, by shifting the resonance away from those of background mechanical modes, we show that this feedback can greatly mitigate degradation due to background resonances. Consequently,  the criteria for achieving quantum squeezing are substantially relaxed, facilitating the entry of existing high-frequency optomechanical devices into the non-classical regime at room temperature. Strikingly, we also establish that spring hardening, shifting the mechanical frequency higher, permits momentum squeezing through continuous position measurement, thus granting optomechanical devices---particularly those operating at low frequencies with large mass---access to momentum below the zero-point motion.  Collectively, these advancements pave the way for room-temperature quantum state preparation and quantum sensing.

\begin{figure}[h]
	\centering
 \includegraphics[width=\columnwidth]{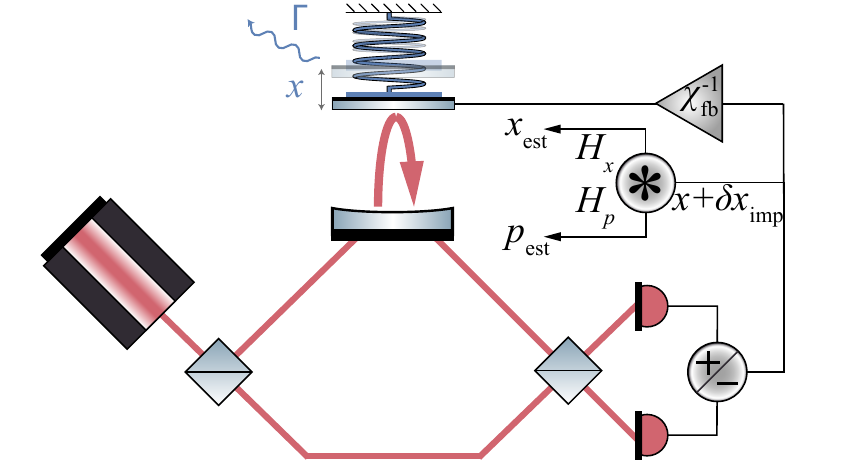}
    
    \caption{\label{fig:diagram} Schematic diagram of state-preparation. A laser resonantly excites an optomechanical cavity. The mechanical position is detected through optical phase quadrature measurement. Feedback is applied to it to control the mechanical resonance frequency. The Wiener filters ($H_x$ and $H_p$) filter the measurement,
	providing optimal estimates of $x$ and $p$.}
\end{figure}

We consider a mechanical oscillator with effective mass $m$ and natural resonance frequency $\Omega_0$, weakly coupled to a thermal bath at temperature $T$. This oscillator undergoes continuous position measurement, and is concurrently manipulated by measurement-based feedback designed to shift its resonance frequency from $\Omega_0$ to $\Omega$. The oscillator's position and momentum, $x$ and $p$, provide a description of its state. They comply with the commutation relation $[x,p] = i \hbar$, fundamentally limiting the precision with which they can be simultaneously known. When the oscillator is cooled to its ground state, its position and momentum are localized to their respective zero-point fluctuations, $x_{\text{zp}}=\sqrt{{\hbar}/{2 m \Omega}}$ and $p_{\text{zp}}=\sqrt{\hbar m \Omega/2}$. For a squeezed state, one observable has uncertainty below the zero-point fluctuations, with the uncertainty in the other correspondingly increased. 

To determine the uncertainty in the observables, we employ state estimation following the approach in Ref.~\cite{meng_mechanical_2020}, but extending it to include feedback control. The estimation relies on the measurement record, with the shared measurement channel servicing both estimation and feedback control, as shown in Fig.~\ref{fig:diagram}. The system features complex dynamics due to the interplay between measurement conditioning, dissipation, bath noise, quantum backaction, and measurement-based feedback force.

While our analysis is applicable more broadly, we focus on implementation using an optomechanical system equipped with an optical cavity, boosting the optomechanical interaction strength~\cite{bowen_quantum_2015,aspelmeyer_cavity_2014}, as shown in Fig.~\ref{fig:diagram}. We adhere to the unresolved sideband regime, for which the optical cavity decay rate $\kappa$ is much larger than both $\Omega_0$ and $\Omega$. In this limit, both the mechanical oscillator and the input field can be adiabatically tracked~\cite{bennett_quantum_2016,doherty_quantum_2012}. The laser frequency is chosen to be on-resonance with the cavity, causing the mechanical position to be imprinted on the output optical phase quadrature~\cite{aspelmeyer_cavity_2014,bowen_quantum_2015}.  Measuring this quadrature provides an inference of the mechanical position,  $x_{\mathrm{meas}}(t)=x(t)+\delta x_{\mathrm{imp}}(t)$, where $\delta x_{\mathrm{imp}}(t)$  is the imprecision noise of the measurement. The optomechanical interaction also introduces measurement backaction, imprinting the intracavity optical amplitude quadrature onto the mechanical resonator.  A feedback force proportional to $x_{\mathrm{meas}}$ is applied to the mechanical oscillator. Experimentally, this could be applied through modulations of the incident optical intensity~\cite{cohadon_cooling_1999,arcizet_radiation-pressure_2006,kleckner_sub-kelvin_2006,meng_optomechanical_2022}, electrical actuation~\cite{poggio_feedback_2007,lee_cooling_2010,harris_minimum_2013,harris_feedback-enhanced_2012,whittle_approaching_2021} or other methods.

We characterize the mechanical motion in the frequency domain by taking the Fourier transform with convention $F(\omega)=\int^{+\infty}_{-\infty}f(t)e^{i \omega t} dt $. The position response is determined by the combined influence of the thermal force $F_{\mathrm{th}}(\omega)$, backaction force $F_{\mathrm{ba}}(\omega)$, and feedback force $F_{\mathrm{fb}}(\omega)$. This is given by~\cite{whittle_approaching_2021}:
\begin{equation}
\label{eq:force}
x(\omega) = \left(F_{\mathrm{th}}(\omega)+F_{\mathrm{ba}}(\omega)+F_{\mathrm{fb}}(\omega)\right) \chi_0(\omega),
\end{equation}
where $\chi_0(\omega)$ is the intrinsic susceptibility defined by $\left. \chi_0^{-1}(\omega)={m}{(\Omega_0^2-\omega^2-i \Gamma \omega)}\right.$ with $\Gamma$ being the mechanical energy damping rate. The feedback force, which depends on the measurement, can be expressed as $\left.F_{\mathrm{fb}}(\omega)=\chi_{\mathrm{fb}}^{-1}(\omega) x_{\mathrm{meas}}(\omega)\right.$. In general, the feedback susceptibility $\chi_{\mathrm{fb}}(\omega)$ is a complex function of frequency, taking into account any filtering of the measured photocurrent, feedback strength and delays in the feedback loop~\footnote{$\chi_{\mathrm{fb}}^{-1}(\omega)=-K m(1-i \omega/ \Gamma_{\mathrm{fb}})e^{i \omega \tau} \approx -K m(1+\omega^2 \tau/ \Gamma_{\mathrm{fb}}+i \omega(\tau -1/  \Gamma_{\mathrm{fb}})) $.}. Here, for simplicity, we choose it to be a real constant, $\left.\chi_{\mathrm{fb}}^{-1}(\omega)=-K m\right.$ with $K$ representing the gain factor of the feedback circuit. The approximation holds for  $\omega \tau \ll 1$~\cite{whittle_approaching_2021}, where $\tau$ is the overall delay in the feedback loop. Choosing $\chi_{\mathrm{fb}}(\omega)$ to be real causes the feedback to induce a mechanical resonance frequency shift.  Solving for $x(\omega)$ under this assumption, Eq.~(\ref{eq:force}) can be re-written as $(F_{\mathrm{th}}(\omega)+F_{\mathrm{ba}}(\omega)+\delta F_{\mathrm{fb}}(\omega))\chi(\omega)=x(\omega)$, where  $\delta F_{\mathrm{fb}}(\omega)=\chi_{\mathrm{fb}}^{-1}(\omega)\delta x_{\mathrm{imp}}(\omega)$ is a force noise term introduced by the measurement imprecision. The modified mechanical susceptibility $\left. \chi^{-1}(\omega)={m} {(\Omega^2-\omega^2-i \Gamma \omega)} \right.$, with shifted mechanical frequency $\Omega=\sqrt{\Omega_0^2+K }$, and unchanged mechanical dissipation rate. 

We now estimate the mechanical state in the presence of both linear position measurement and feedback control. The measurement signal satisfies the commutation relation $[x_{\mathrm{meas}}(t),x_{\mathrm{meas}}(t')] = 0$~\cite{buonanno_signal_2002,chen_causal_2023}. This enables it to be treated classically and classical filtering to be used to estimate the mechanical position and momentum. The estimates are given by $\left.x_\text{est}(t) = H_{x}(t) \circledast x_{\mathrm{mea}}(t)\right.$ and $\left. p_\text{est}(t) = H_{p}(t) \circledast x_{\mathrm{mea}}(t)\right.$, where $H_{x}(t)$ and $H_{p}(t)$ are filter functions. The optimal estimates, minimizing the mean-squared error, are provided by the causal Wiener filters (see derivation in Ref.\footnotetext[100]{Supplemental material}~\footnotemark[100])~\cite{wiener_extrapolation_1964}
\begin{equation}
\label{eq:filter}
{H}_{o}(\omega) =\left(A_{o}-i B_{o} \omega\right) \chi(\omega)^{\prime}, \quad o \in \{x, p\} 
\end{equation}
where $A_{o}$ and $B_{o}$ are frequency-independent coefficients given in Ref.~\footnotemark[100], and $\left.\chi(\omega)^{\prime} ={1}/ {(\Omega^{\prime 2}-\omega^2-i \Gamma^{\prime} \omega)}\right.$ corresponds mathematically to a modified mechanical susceptibility that peaks at $\left.\Omega^{\prime} =\left(\Omega_0^4+\Omega_{\mathrm{meas}}^4\right)^{1 / 4}\right.$ and has linewidth $\left.\Gamma^{\prime} =\sqrt{\Gamma^2-2 \Omega_0^2+2 \Omega^{\prime 2}}\right.$. The frequency $\Omega_{\mathrm{meas}}$ is an important parameter that emerges from the model and is independent of feedback control. We term this the \textit{characteristic measurement frequency}. It is given by $\left. \Omega_{\mathrm{meas}}=2 (\eta C n_{\mathrm{tot}})^{1/4}\sqrt{ \Gamma \Omega_0} \right.$, where $\eta$ is the detection efficiency, $\left.n_{\mathrm{tot}}=n_{\mathrm{th}}+C+1/2	\right.$ is the intrinsic total occupancy (including the 1/2 quanta of vacuum energy for succinctness), $n_{\mathrm{th}}\approx k_B T / \hbar \Omega$ is the average thermal phonon occupation~\cite{kohen_phase_1997}, and $C=4g^2/\kappa \Gamma$ is the optomechanical cooperativity with $g$ the coherent amplitude boosted optomechanical coupling rate of the un-frequency-shifted oscillator~\cite{aspelmeyer_cavity_2014,bowen_quantum_2015}. For an oscillator without feedback control, $\left.\Omega_{\mathrm{meas}}>\Omega_0\right.$ defines the regime in which the rotating wave approximation (RWA) is invalid~\cite{meng_mechanical_2020}.

As the measurement is linear, the oscillator's conditional state can be fully described by its covariance matrix, $\mathbb{V}=\left(\begin{array}{cc}
V_{\delta X \delta X }  & V_{\delta X \delta P } \\
V_{\delta X \delta P } & V_{\delta P \delta P }
\end{array}\right),$ 
where here and throughout we normalize the variances and covariances by their respective zero-point motion (\textit{i.e.}, $\left. V_{\delta X \delta X }=V_{\delta x \delta x }/x_{\mathrm{zp}}^2 \right.$, $\left.V_{\delta P \delta P }=V_{\delta p \delta p }/p_{\mathrm{zp}}^2\right.$, $\left.V_{\delta X \delta P }=V_{\delta x \delta p }/x_{\mathrm{zp}} p_{\mathrm{zp}}\right.$, and $\left.{\delta x, \delta p} = {x- x_{\mathrm{est}}, p - p_{\mathrm{est}}}\right.$). Employing the optimal Wiener filters, we derive analytical expressions for the covariance matrix. These are characterized by five dimensionless variables: $\eta$, $C$, $n_{\mathrm{th}}$, $Q$, and the relative frequency ratio $R=\Omega/\Omega_0$. In the general case, the expressions are non-intuitive. For completeness, we include them in Ref.~\footnotemark[100].

\begin{figure}[h]
	\centering
    \includegraphics[width=\columnwidth]{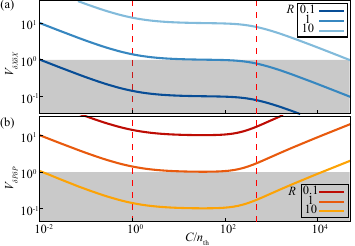}   
	\caption{\label{fig:line_curve} Characterization of conditional states under varying feedback with $n_{\mathrm{th}}=2 \times 10^3$ and $\eta=1$ as a function of $C$. (a)~Position and (b)~momentum  variances for several $R$,  are shown with cold and warm colors, respectively. Grey area: squeezing regime. Red dashed lines: backaction dominated  ($C>n_{\mathrm{th}}$) and quantum squeezing boundaries ($C>Q/2$) without feedback.}
\end{figure}

Fig.~\ref{fig:line_curve} plots the diagonal terms of the covariance matrix as a function of $C$ for different choices of feedback gain. In the absence of feedback control, both mechanical position and momentum variances decrease as $C$ increases in the thermal noise dominated regime ($C \ll n_{\mathrm{th}}$). The variances stabilize, becoming equally localized close to the zero-point fluctuations in the backaction-dominated regime ($C \gg n_{\mathrm{th}}$). Further increasing $C$ into the squeezing regime ($C > Q/2$~\footnote{Inside the backaction dominated regime, the squeezing criterion $C>\left(n_{\mathrm{tot}} Q^2/ 4\right)^{1 / 3}$ can be simplified to $\left.C > Q/2\right.$}) leads to a reduction in position variance below the zero-point fluctuations, while the momentum variance increases, consistent with predictions in previous work~\cite{meng_mechanical_2020}. 

Feedback control can influence the conditional state without altering the intrinsic backaction and thermal noise. As shown in Fig.~\ref{fig:line_curve}, in the softening scenario ($R<1$) we find that the position localization is enhanced and the momentum localization is degraded. As $R$ decreases, the conditional position variance can be reduced beneath the zero-point fluctuations even in scenarios where the no-feedback quantum squeezing criterion defined in Ref.~\cite{meng_mechanical_2020} is not satisfied. In complement to this, the hardening scenario ($R>1$) weakens position localization, making quantum squeezing more challenging to achieve. Surprisingly, given the use of position measurement, this hardening allows momentum squeezing. This is noteworthy given that momentum rather than position squeezing is required for quantum enhanced force sensing~\cite{khosla_quantum_2017}. The conditional momentum variance initially decreases with increasing $C$ and eventually increases for sufficient high $C$. Thus, unlike position squeezing, which monotonically improves as  $C$ increases, we find that there is an optimal $C$ for maximal momentum squeezing.

\begin{figure}[h]
	\centering \includegraphics[width=\columnwidth]{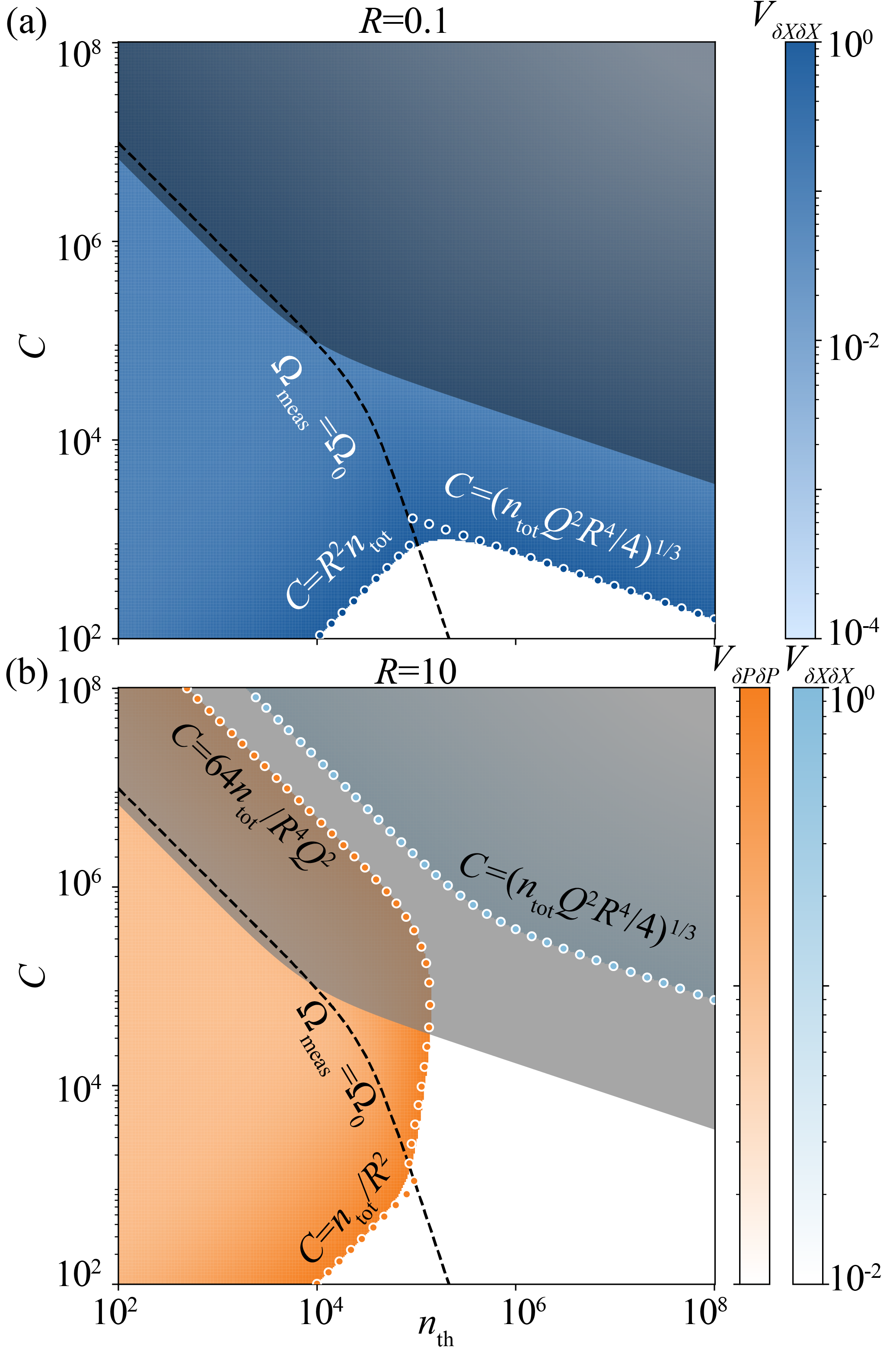}   
	\caption{\label{fig:compare_feedback} Characterization of the conditional variances as a function of $n_{\mathrm{th}}$ and $C$ for $R=0.1$~(a) and $R=10$~(b), with $\eta=1$, at room temperature. Black dashed lines: RWA breakdown criterion. Dotted lines in corresponding colors:  respective squeezing thresholds. Grey shaded regions: position squeezing regimes when $\left.R=1\right.$. Blue shading: position squeezing. Orange shading: momentum squeezing. }
\end{figure}

To provide a comprehensive understanding of how the feedback control affects the squeezed state in different measurement regimes, we characterize the conditional position and momentum variances as a function of $n_{\mathrm{th}}$ and $C$ in Fig.~\ref{fig:compare_feedback}. 

Fig.~\ref{fig:compare_feedback}~(a) illustrates the position variance within the softening scenario for $R=0.1$. Importantly, the parameter regime in which quantum squeezing occurs significantly expands when feedback is applied (blue \textit{c.f.} grey shaded regions), enabling squeezing in regimes previously deemed inaccessible. This eases the requirements for achieving quantum squeezing, opening the possibility for quantum state preparation with a significantly wider class of optomechanical systems.

Fig.~\ref{fig:compare_feedback}~(b) presents position (blue) and momentum (orange) variances within the hardening regime for $R=10$. Notably, the parameter regime supporting position squeezing contracts compared to the case without feedback.  However, this contraction is accompanied by the emergence of a new regime characterized by momentum squeezing as shown in the orange region in Fig.~\ref{fig:compare_feedback}~(b).

To further understand the impact of feedback on the conditional state and its subsequent influence on the criteria for quantum squeezing, we simplify the expressions governing the conditional state by taking the limit of a high $Q$ oscillator ($Q\gg1$) and finite measurement strength, \textit{i.e.}, $C \not\to\ 0$.  The covariance matrix of the conditional state can then be greatly simplified to
\begin{equation}
\label{eq: covariance_matrix}
\mathbb{V}= \frac{2 n_{\text {tot }}\Gamma }{\Omega_{\text {meas }}^4}\left(\begin{array}{cc}
	\Gamma^{\prime} R \Omega_0^2 & {2 \Omega_0}\left(\Omega^{\prime 2}-\Omega_0^2\right) \\
{2 \Omega_0}\left(\Omega^{\prime 2}-\Omega_0^2\right) & \Gamma^{\prime}\Omega^{\prime 2} / R
\end{array}\right).
\end{equation}
The purity of the system $\mathcal{P}=\sqrt{{\eta  C} / {n_{\text{tot}}}}$ is independent of feedback control and is always equal to or below unity, ensuring the state's physical validity. The diagonal terms in Eq.~(\ref{eq: covariance_matrix}) characterize the position and momentum squeezing. The non-zero off-diagonal covariance terms indicate that the position and momentum are correlated so that the optimal squeezing occurs at an in-between quadrature (detailed in Ref.~\footnotemark[100]). However, for typical measurement regimes, the variance of this quadrature is close to the minimum of $V_{\delta x \delta x}$ and $V_{\delta p \delta p}$. We thus confine our analysis to position and momentum.

We establish intuitive criteria for position and momentum squeezing by simplifying Eq.~(\ref{eq: covariance_matrix}) and determining when its diagonal elements are less than unity. To do this, we consider two distinct scenarios: the non-RWA regime $\Omega_{\text{meas}} \gg \Omega_0$, for which $\Omega' \rightarrow \Omega_{\text{meas}}$ and $\Gamma' \rightarrow \sqrt{2}\Omega_{\text{meas}}$; and the RWA regime  $\Omega_{\text{meas}} \ll \Omega_0$, for which $\Omega' \rightarrow \Omega_0$ and $\Gamma' \rightarrow \Omega_{\text{meas}}^2/\Omega$. To facilitate comparison with optomechanical systems that operate without feedback control, we formulate these criteria using unshifted dimensionless parameters $\eta$, $C$, $n_{\mathrm{th}}$ and  $Q$,  consistent with extant literature~\cite{meng_mechanical_2020}.

In the non-RWA regime, we find that position squeezing is achieved when $ C>\left(n_{\mathrm{tot}} Q^2 R^4 / 4\right)^{1 / 3}/\eta$. This criterion can be further simplified if either thermal noise ($n_{\mathrm{th}} \gg C$) or backaction noise ($C \gg n_{\mathrm{th}}$) dominates the noise driving the mechanical oscillator. In the former case, we find that squeezing occurs when $\left.C>\left(n_{\mathrm{th}} Q^2 R^4 / 4\right)^{1 / 3}/\eta \right.$, while in the latter it occurs when $\left.C>Q R^2 /2 \eta^{3/2}\right.$. Hence, feedback reduces the required $C$ for squeezing by factors of $R^{4/3}$ and $R^2$, respectively. In an ideal scenario with perfect detection efficiency ($\left.\eta = 1\right.$) and in the backaction dominated regime, the position squeezing criterion reduces to $\left.\Omega_{\mathrm{meas}}>\sqrt{2} R \Omega_0\right.$, so that position squeezing is achievable when the measurement bandwidth surpasses the feedback-shifted mechanical frequency by a factor of $\sqrt{2}$. 

Within the RWA regime, our findings reveal that squeezing can be achieved when $C>n_{\mathrm{tot}}R^2/\eta$, independent of whether the system is in the backaction or thermal noise dominated regimes.

Analogous to position squeezing, we establish the momentum squeezing criteria as $ C>64 n_{\mathrm{tot}}^3/\eta R^4 Q^2$ and $ C>n_{\mathrm{tot}}/\eta R^2$ outside and inside the RWA regimes, respectively. Notably, feedback not only makes momentum squeezing possible, but can also be used to tune the requirements for squeezing through adjustable feedback strength.

The position and momentum squeezing criteria, deduced in the previous paragraphs are depicted by blue and orange dotted lines in Fig.~\ref{fig:compare_feedback} panels (a) and (b), agreeing well with the full model.

\begin{figure}[h] \includegraphics[width=\columnwidth]{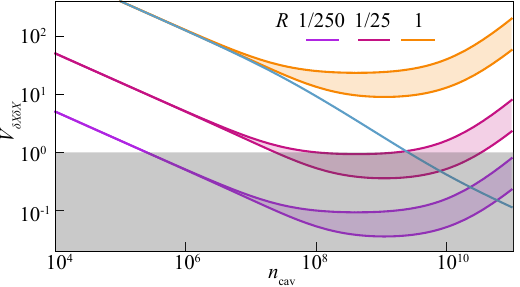}\caption{\label{fig:square_membrane_feedback}Conditional position variance as a function of $n_{\mathrm{cav}}$ for different $R$ at room temperature, with (colored lines) and without (black line) second-order mode noise. The split-colored lines represent the lower and upper bounds of the conditional variance when taking the simplified model. Fundamental mode optomechanical parameters: $G/2\pi\sim$14~MHz/nm,  $\Omega_0/2\pi\sim$0.8 MHz, $Q\sim~10^6$ and $m\sim$7~ng. Second mode: $ G_2 = G$, $\Omega_2 = 2\Omega_0$, $Q_2 = Q$, and $m_2 = m$. $\eta$=0.63~\cite{purdy_observation_2013}.}
\end{figure}

To evaluate the experimental feasibility of our proposed method, we consider the membrane-in-the-middle optomechanical system of Ref.~\cite{purdy_observation_2013}. This square membrane has optomechanical coupling strength $G/2\pi\sim$14~MHz/nm, detection efficiency $\eta$=0.63, and a fundamental mode with $\Omega_0/2\pi\sim$0.8~MHz, $Q\sim~10^6$ and $m\sim$7 ng~\cite{purdy_observation_2013}. With these parameters, at room temperature position squeezing can be achieved with $3\times10^9$ intracavity photons. With feedback to achieve $R=1/10$, we find this to reduce to  $9\times10^7$ intracavity photons. By contrast, increasing the resonance frequency by a factor of ten does not enable momentum squeezing. Momentum squeezing is possible though, for higher $R$. For $R=20$, it can be achieved with $3\times10^8$ intracavity photons. Ref.~\cite{purdy_observation_2013} used $4\times 10^8$ intracavity photons, indicating that feedback could realistically enable both position and momentum squeezing.

Background mechanical resonances have been identified to significantly degrade measurement-based state preparation in a recent experiment~\cite{meng_measurement-based_2022}. It is interesting to examine if feedback mechanisms can counteract this effect and facilitate mechanical squeezing. To explore this, we examine the influence of the second higher-order mode of the membrane considered above. We use the parameters $\Omega_2 = 2\Omega_0$, $Q_2 = Q$, $m_2 = m$, and a conservative estimate for optomechanical coupling strength of $ G_2 = G$~\cite{purdy_observation_2013, wilson_cavity_2012}. The noise of the second resonance peak has a flat spectrum at frequencies well below $\Omega_2$ and is suppressed by the Wiener filter at frequencies well above $\Omega_0$~\cite{meng_measurement-based_2022}. As such, it can be well-approximated as an additional white noise source. This simplifies the calculation of the conditional variance, allowing analytical solutions. The amplitude of the additional noise is bounded from below by the noise from the second mode at zero frequency, and from above by the maximum noise from the second mode across the full frequency band that the Wiener filter samples. This results in lower and upper bounds, respectively, on the conditional variance, as further justified in Ref.~\footnotemark[100].

Contrary to the case of a single mechanical mode, for which the position variance decreases monotonically with measurement strength~\cite{meng_mechanical_2020}, within the validity of the white noise approximation our analysis shows that the presence of a second mode introduces an optimal measurement strength, above which the variance degrades. This is shown in Fig.~\ref{fig:square_membrane_feedback}. The optimum measurement is independent of feedback, and for the chosen parameters, this optimum occurs around \(  10^9 \)  intracavity photons. As shown in Fig.~\ref{fig:square_membrane_feedback}, even at the optimum intracavity photon number, squeezing is not possible in the absence of feedback. However, it is possible with feedback for $R\lesssim 1/25$.

In conclusion, we have shown that feedback control can substantially relax the requirements for measurement-based mechanical position squeezing, and that it can enable momentum squeezing. Furthermore, we find that feedback control can suppress degradation in squeezing due to the presence of background mechanical resonances. This paves the way for room-temperature quantum state preparation and sensing.

%TC:ignore
\begin{acknowledgments}
	We would like to thank J.~X. Chen, S.~A. Fedorov, J.~S. Bennett, A. Schliesser,  K. M\o lmer, E. Zeuthen, E. S. Polzik  for useful discussion. This work was supported by the Australian Research Council Centres of Excellence for Engineered Quantum Systems (EQUS, CE170100009) and Quantum Biotechnology (CE230100021), Air Force Office of Scientific Research under award number FA9550-20-1-0391 and VILLUM FONDEN under a Villum Investigator Grant No. 25880. C.M. acknowledges funding from the European Union's Horizon Europe research and innovation programme under the Marie Sk\l{}odowska-Curie grant agreement No. 101110196.
\end{acknowledgments}
%TC:endignore


% %\nocite{*}
% %TC:ignore
% \section*{References}
% %\nocite{*}
% %\footnotesize \renewcommand{\refname}{\vspace*{-30pt}}
% \bibliographystyle{apsrev4-2} % Tell bibtex which bibliography style to use
% \bibliography{Ref} % Tell bibtex which .bib file to use
% %TC:endignore

\clearpage

\pagebreak

\widetext
\begin{center}
	\textbf{\large Supplemental Material: Enhancement of mechanical squeezing via feedback control}
\end{center}
%%%%%%%%%% Merge with supplemental materials %%%%%%%%%%
%%%%%%%%%% Prefix a "S" to all equations, figures, tables and reset the counter %%%%%%%%%%
\setcounter{equation}{0}
\setcounter{figure}{0}
\setcounter{table}{0}
\setcounter{page}{1}
\makeatletter
\renewcommand{\theequation}{S\arabic{equation}}
\renewcommand{\thefigure}{S\arabic{figure}}

\section{Model of measurement and feedback}

Estimating the state utilizing optimal Wiener filters necessitates a comprehensive system model. This model is effectively represented through the spectral densities of displacement and force noise. The displacement spectral density is formulated as~\cite{whittle_approaching_2021}
\begin{equation}
S_{x x}(\omega)=(S_{F F}^{\text {tot }}(\omega)+S_{F F}^{\mathrm{\delta f b}}(\omega))|\chi(\omega)|^2,
\end{equation}
where $S_{FF}^{\mathrm{tot}}(\omega)$ encapsulates the total spectral density of both thermal $S_{FF}^{\mathrm{th}}(\omega)$ and backaction $S_{FF}^{\mathrm{ba}}(\omega)$ forces. This is expressed as
\begin{equation}
S_{FF}^{\mathrm{tot}}(\omega) = 2n_{\mathrm{tot}} S_{xx}^{\mathrm{zp}}(\Omega_0) /| \chi_0(\Omega_0)|^2,
\end{equation}
where $S_{xx}^{\mathrm{zp}}(\Omega_0) = 2 x_{\mathrm{zp}}(\Omega_0)^2/\Gamma$ defines the intrinsic zero-point noise of the oscillator~\cite{clerk_introduction_2010,whittle_approaching_2021}. The spectral density characterizing the imprecision feedback force given by $S_{FF}^{\mathrm{\delta fb}}(\omega)=S_{x x}^{\mathrm{imp}}(\omega)|\chi_{\mathrm{fb}}(\omega)|^{-2}$, where $\left.S_{x x}^{\mathrm{imp}}(\omega)={S_{xx}^{\mathrm{zp}}(\Omega_0)}/{8 \eta C}\right.$ indicates the measurement-imprecision noise at the optical vacuum noise level.

Subsequently, the noise equivalent Power Spectral Density (PSD) of the measurement laser phase quadrature is given by~\cite{clerk_introduction_2010,bowen_quantum_2015,whittle_approaching_2021}
\begin{equation}
S_{x x}^{\mathrm{meas}}(\omega)=S_{x x}(\omega)+S_{x x}^{\mathrm{imp}}(\omega)+2S_{x \delta x_{\mathrm{imp}}}(\omega),
\end{equation}
where $S_{x \delta x_{\mathrm{imp}}}(\omega)$ represents the cross-correlation between the displacement and imprecision noise introduced by the feedback. The cross-spectral density is defined by~\cite{clerk_introduction_2010,bowen_quantum_2015}
\[
S_{AB}(\omega)=\int^\infty_{-\infty}e^{i\omega t}\langle A(t)B(0)\rangle \text{d}t=\int^{\infty}_{-\infty} \frac{d\omega^\prime}{2\pi}\langle{A^{\dag}(-\omega)}{B(\omega\prime)}\rangle.\]

\section{Measurement bandwidth}

In the high $Q$ limit where $Q=\Omega_0/\Gamma\gg1$, $\Omega_{\text{meas}}$ can be directly related to the measurement bandwidth $B_{\text{meas}}$ for which the intrinsic thermal noise is resolved above the optical shot noise. Inside the RWA regime, $\Omega_{\mathrm{meas}}=\sqrt{B_{\text{meas}}\Omega_0}$, while outside this regime $\left.\Omega_{\mathrm{meas}} = B_{\text{meas}}\right.$. 

\section{Filter functions}

The causal Wiener filter is \cite{wiener_extrapolation_1964}
\begin{equation}
\label{eqns:filter1}
H_{o}(\omega)=\frac{1}{M_{x}^{\mathrm{meas}}(\omega)}\left[\frac{S_{o x^{\mathrm{meas}}}(\omega)}{{M_{x}^{\mathrm{meas}}(\omega)^*}}\right]_{+}, \quad o \in \{x, p\} 
\end{equation}
where  $S_{ox^{\mathrm{meas}}}(\omega)=S_{o x}(\omega)+S_{o \delta x_{\mathrm{imp}}}(\omega)$, and $M_{x}^{\mathrm{meas}}(\omega)$ is the causal spectral factor that satisfies $\left.S_{xx}^{\mathrm{meas}}(\omega)=M_{x}^{\mathrm{meas}}(\omega)M_{x}^{\mathrm{meas}(\omega)^*}\right.$ and only has poles and zeros in the lower half of the complex plane.  $[...]_+$ denotes the causal part of the function. In general, any function may be separated into the sum of its causal and anti-causal parts \cite{wiener_extrapolation_1964}, which can be decomposed by factorizing the poles of the denominator into the upper and the lower halves of the complex plane and finding the partial fraction decomposition. Expanding Eq.~(\ref{eqns:filter1}) using this procedure, we find the coefficients of the filter functions in Eq.~(2) are
\begin{equation}
\begin{aligned}
& A_x=\frac{\Omega_{\text {meas }}^4\left(\Gamma^2+\Gamma \Gamma^{\prime}+\Omega^{\prime 2}-\Omega^2\right)+\left(\Omega_0^2-\Omega^2\right)\left(\Omega_0^2\left(\Gamma\left(\Gamma+\Gamma^{\prime}\right)+\Omega^{\prime 2}\right)-\left(\Omega_0^2+\Omega^{\prime 2}\right) \Omega^2+\Omega^4\right)}{\left(\Gamma\left(\Gamma+\Gamma^{\prime}\right) \Omega^{\prime 2}+\Omega^{\prime 4}+\left(\Gamma^{\prime}\left(\Gamma+\Gamma^{\prime}\right)-2 \Omega^{\prime 2}\right) \Omega^2+\Omega^4\right)}, \\
& B_x=\frac{\Omega_{\text {meas }}^4\left(\Gamma+\Gamma^{\prime}\right)+\left(\Omega_0^2-\Omega^2\right)\left(\Gamma\left(\Omega_0-\Omega^{\prime}\right)\left(\Omega_0+\Omega^{\prime}\right)+\Gamma^{\prime}\left(\Omega_0^2-\Omega^2\right)\right)}{\left(\Gamma\left(\Gamma+\Gamma^{\prime}\right) \Omega^{\prime 2}+\Omega^{\prime 4}+\left(\Gamma^{\prime}\left(\Gamma+\Gamma^{\prime}\right)-2 \Omega^{\prime 2}\right) \Omega^2+\Omega^4\right)}, \\
& A_p=m \frac{\Omega^2\left(-\Omega_{\text {meas }}^4\left(\Gamma+\Gamma^{\prime}\right)-\left(\Omega_0^2-\Omega^2\right)\left(\Gamma\left(\Omega_0^2-\Omega^{\prime 2}\right)+\Gamma^{\prime}\left(\Omega_0^2-\Omega^2\right)\right)\right)}{\left(\Gamma\left(\Gamma+\Gamma^{\prime}\right) \Omega^{\prime 2}+\Omega^{\prime 4}+\left(\Gamma^{\prime}\left(\Gamma+\Gamma^{\prime}\right)-2 \Omega^{\prime 2}\right) \Omega^2+\Omega^4\right)}, \\
& B_p=m \frac{\Omega_{\text {meas }}^4\left(\Omega^{\prime 2}-\Omega^2\right)+\left(\Omega_0^2-\Omega^2\right)\left(\Gamma^2 \Omega^{\prime 2}+\Gamma \Gamma^{\prime} \Omega^2+\left(\Omega^{\prime 2}-\Omega^2\right)\left(\Omega_0^2-\Omega^2\right)\right)}{\left(\Gamma\left(\Gamma+\Gamma^{\prime}\right) \Omega^{\prime 2}+\Omega^{\prime 4}+\left(\Gamma^{\prime}\left(\Gamma+\Gamma^{\prime}\right)-2 \Omega^{\prime 2}\right) \Omega^2+\Omega^4\right)}.
\end{aligned}
\end{equation}

\section{Covariance matrix}

Employing the optimal filters in Eq.~(2), we derive the variances and covariance using~\cite{bowen_quantum_2015}
\begin{equation}
\begin{aligned}
& V_{\delta x \delta x}=\int^{\infty}_{-\infty}\frac{d\omega}{2\pi}S_{\delta x \delta x}, \\
& V_{\delta p \delta p}=\int^{\infty}_{-\infty}\frac{d\omega}{2\pi}S_{\delta p \delta p}, \\
& V_{\delta x \delta p}=\int^{\infty}_{-\infty}\frac{d\omega}{2\pi}\mathrm{Re}\{S_{\delta x \delta p}\}.
\end{aligned}
\end{equation}
The expressions for the normalized, fully analytical variances and covariance of conditional position and momentum are
\begin{equation}
	\label{eq:full_covariance_matrix}
	\begin{aligned}
	& V_{\delta X \delta X}=\frac{1}{8 C R \Gamma^2} \left(16 C n_{\text {tot }} \Gamma^2+\left(-1+R^2\right)^2 \Omega_0^2+\left(\Omega_0^ { 2 } \left(-2 R^2\left(-1+R^2\right) \Omega_0^2\left(16 C n_{\text {tot }} \Gamma^2-\left(-1+R^2\right) \Omega_0^2\right)\left(R^4 \Omega_0^4+\right.\right.\right.\right. \\
	& \left.2 \Gamma\left(\Gamma+\Gamma^{\prime}\right) \Omega^{\prime 2}+\Omega^{\prime 4}+R^2 \Omega_0^2\left(-\Gamma^2+\Gamma^{\prime 2}-2 \Omega^{\prime 2}\right)\right)- R^2\left(-1+R^2\right)^2 \Omega_0^2(R^6 \Omega_0^6+\Gamma^2 \Omega^{\prime 4}+R^4 \Omega_0^4\left(\Gamma^{\prime 2}-2 \Omega^{\prime 2}\right)+
	\\
	&R^2 \Omega_0^2 \Omega^{\prime 2}\left(2 \Gamma \Gamma 1+\Omega^{\prime 2}\right))- \left(-16 C n_{\text {tot }} \Gamma^2+\left(-1+R^2\right) \Omega_0^2\right)^2 \left.\left.\left(R^4 \Omega_0^4+R^2 \Omega_0^2\left(-\Gamma^2+\Gamma^{\prime 2}-2 \Omega^{\prime 2}\right)+\left(\Gamma\left(\Gamma+\Gamma^{\prime}\right)+\Omega^{\prime 2}\right)^2\right)\right)\right) / \\
	& \left.\left(R^4 \Omega_0^4+\Gamma\left(\Gamma+\Gamma^{\prime}\right) \Omega^{\prime 2}+\Omega^{\prime 4}+R^2 \Omega_0^2\left(\Gamma 1^{\prime}\left(\Gamma+\Gamma^{\prime}\right)-2 \Omega^{\prime 2}\right)\right)^2\right) \\
	\\
	& V_{\delta P \delta P}=\left(16 C n_ { \text{tot}} \Gamma ^ { 2 } \left(-2 R^2\left(\Gamma+\Gamma^{\prime}\right)^2 \Omega_0^6-R^8 \Omega_0^8+\right.\right. R^4 \Omega_0^4\left(\Gamma^{\prime 2}\left(\Gamma+\Gamma^{\prime}\right)^2+2\left(\Gamma^2+4 \Gamma \Gamma^{\prime}+2 \Gamma^{\prime 2}\right) \Omega_0^2-2 \Omega_0^4\right)+R^6(-2 \Gamma \Gamma^{\prime} \Omega_0^6+
	\\
	&
	4 \Omega_0^8)+  2 R^2 \Omega_0^2\left(\Gamma \Gamma^{\prime}\left(\Gamma+\Gamma^{\prime}\right)^2-\left(\left(-2+R^2\right) \Gamma^2+R^2 \Gamma \Gamma^{\prime}+2 R^2 \Gamma^{\prime 2}\right) \Omega_0^2+2\left(1-2 R^2\right) \Omega_0^4\right) \Omega^{\prime 2}+  (\Gamma^2\left(\Gamma+\Gamma^{\prime}\right)^2-2(\left(1+R^2\right) \Gamma^2
	\\
	&
	+R^2 \Gamma \Gamma^{\prime}-R^2 \Gamma^{\prime 2}) \Omega_0^2+2\left(-1+2\left(R^2+R^4\right)\right) \Omega_0^4) \Omega^{\prime 4}+ \left.2\left(\Gamma\left(\Gamma+\Gamma^{\prime}\right)-2 R^2 \Omega_0^2\right) \Omega^{\prime 6}+\Omega^{\prime 8}\right)- \\
	& 256 C^2 n_{\text {tot }}^2 \Gamma^4 \Omega_0^2\left(R^4 \Omega_0^4+\Omega^{\prime 4}+R^2 \Omega_0^2\left(\left(\Gamma+\Gamma^{\prime}\right)^2-2 \Omega^{\prime 2}\right)\right)+ \left(-1+R^2\right)^2 \Omega_0^2\left(\Omega^{\prime 4}\left(\Gamma \Gamma^{\prime}-\Omega_0^2+\Omega^{\prime 2}\right)\left(\Gamma\left(2 \Gamma+\Gamma^{\prime}\right)+\Omega_0^2+\Omega^{\prime 2}\right)+\right. \\
	& R^2 \Omega_0^2\left(-\left(\Gamma+\Gamma^{\prime}\right)^2 \Omega_0^4+2\left(\Gamma \Gamma^{\prime 3}+\Omega_0^4+2 \Gamma^2\left(\Gamma^{\prime 2}+\Omega_0^2\right)\right) \Omega^{\prime 2}+\right. \left.\left(-3 \Gamma^2-2 \Gamma \Gamma^{\prime}+2\left(\Gamma^{\prime 2}+\Omega_0^2\right)\right) \Omega^{\prime 4}-4 \Omega^{\prime 6}\right)
	\\
	&
	+R^6 \Omega_0^6\left(\Gamma^{\prime 2}+2\left(\Omega_0-\Omega^{\prime}\right)\left(\Omega_0+\Omega^{\prime}\right)\right)+ \left.\left.R^4 \Omega_0^4\left(\Gamma^{\prime 4}+2 \Gamma^{\prime 2} \Omega_0^2-\Omega_0^4-4\left(\Gamma^{\prime 2}+\Omega_0^2\right) \Omega^{\prime 2}+5 \Omega^{\prime 4}+2 \Gamma \Gamma^{\prime}\left(\Gamma^{\prime 2}+2 \Omega_0^2-\Omega^{\prime 2}\right)\right)\right)\right) / \\
	& \left(8 C R \Gamma^2\left(R^4 \Omega_0^4+\Gamma\left(\Gamma+\Gamma^{\prime}\right) \Omega^{\prime 2}+\Omega^{\prime 4}+R^2 \Omega_0^2\left(\Gamma^{\prime}\left(\Gamma+\Gamma^{\prime}\right)-2 \Omega^{\prime 2}\right)\right)^2\right) \\
	\\
	& V_{\delta X \delta P}=\frac{\Omega_0^3\left(16 C n_{\text {tot }} \Gamma^2\left(\Gamma+\Gamma^{\prime}\right)+(-1+R)(1+R)\left(-\left(\left(\Gamma+\Gamma^{\prime}-R^2 \Gamma^{\prime}\right) \Omega_0^2\right)+\Gamma \Omega^{\prime 2}\right)\right)^2}{8 C \Gamma\left(R^4 \Omega_0^4+\Gamma\left(\Gamma+\Gamma^{\prime}\right) \Omega^{\prime 2}+\Omega^{\prime 4}+R^2 \Omega_0^2\left(\Gamma^{\prime}\left(\Gamma+\Gamma^{\prime}\right)-2 \Omega^{\prime 2}\right)\right)^2}. \\
	&
	\end{aligned}
\end{equation}

In the main text of our analysis, Eq.~(\ref{eq:full_covariance_matrix}) has been simplified to Eq.~(3) by considering a high $Q$ oscillator. Further simplification of Eq.~(3) is achievable, both within and outside the RWA regime.

Far outside the RWA regime ($\Omega_{\mathrm{meas}} \gg \Omega_0$), the covariance matrix derived in Eq.~(3) can then be simplified to
\begin{equation}
\label{eq:nonRWA_covariance_matrix}
\mathbb{V}_{\Omega_{\mathrm{meas}}\gg \Omega_0}=\frac{1}{\mathcal{P}}\left(\begin{array}{cc}
\sqrt{2} \Omega / \Omega_{\text {meas }} & 1 \\
1 & \sqrt{2} \Omega_{\text {meas }} / \Omega
\end{array}\right).
\end{equation}

Conversely, deeply inside the RWA regime ($\Omega_{\mathrm{meas}} \ll \Omega_0$), yet outside the weak measurement regime~\footnote{The protocol does not accommodate the weak measurement regime, as the thermal motion cannot be resolved from the measurement, thereby rendering feedback application unfeasible. Furthermore, it is worth noting that, through feedback control, shifting $\Omega$ down nearly to DC  only requires $\Omega_{\mathrm{meas}} \gg \sqrt{ \Omega_0 \Gamma}$ while ground state cooling requires $\Omega_{\mathrm{meas}} \gg 2 n_{\mathrm{th}}n_{\mathrm{tot}}\sqrt{ \Omega_0 \Gamma}$, \textit{i.e.}, $\eta C\gg n_{\mathrm{th}}$. Therefore, feedback softening does not have as significant a technical hurdle as feedback cooling.} ($\left.\Omega_{\mathrm{meas}} \gg \sqrt{ \Omega_0 \Gamma}\right.$)~\cite{bowen_quantum_2015}, the correlation is suppressed so that in this regime $\theta=0$ and the covariance matrix simplifies to
\begin{equation}
	\label{eq:RWA_covariance_matrix}
	\mathbb{V}_{\Omega_{\mathrm{meas}}\ll \Omega}=\frac{1}{\mathcal{P}}\left(\begin{array}{cc}
		R & 0 \\
	0 & R^{-1}
	\end{array}\right).
\end{equation}

\section*{Minimum variance}

The minimum variance $V_{\delta X^{\theta} \delta X^{\theta}}$ of the optimal quadrature $\left. \delta X^{\theta}=\delta X \mathrm{cos} \theta+ \delta P \mathrm{sin} \theta \right.$ occurs for an angle $\left. \theta=-\mathrm{arctan}(\sqrt{2}/(\Omega_{\mathrm{meas}}/\Omega-\Omega/\Omega_{\mathrm{meas}}))/2\right.$. This minimum variance can be expressed in terms of the covariance matrix as~\cite{meng_mechanical_2020}
\begin{equation}
\label{eq:V_min}
	V_{\delta X^{\theta} \delta X^{\theta}} = \frac{V_{\delta X \delta X} + V_{\delta P \delta P} - \sqrt{V_{\delta X \delta X}^2 + V_{\delta P \delta P}^2 + 4V_{\delta X \delta P}^2 - 2V_{\delta X \delta X} V_{\delta P \delta P} }}{2}.
\end{equation} 
By substituting Eq.~(\ref{eq:nonRWA_covariance_matrix}) into Eq.~(\ref{eq:V_min}), we derive the simplified form
\begin{equation}
\label{eq:V_min_sim}
	V_{\delta X^{\theta} \delta X^{\theta}} = \frac{V_{\delta X \delta X} + V_{\delta P \delta P} - \sqrt{V_{\delta X \delta X}^2 + V_{\delta P \delta P}^2}}{2}.
\end{equation} 
In the regime of a strongly squeezed state, where $\left. \mathrm{min}\{V_{\delta X \delta X}, V_{\delta P \delta P}\} \ll 1 \right.$, a first-order Taylor expansion applied to Eq.~(\ref{eq:V_min_sim}) yields:
\begin{equation}
	V_{\delta X^{\theta} \delta X^{\theta}} = \frac{\mathrm{min}\{V_{\delta X \delta X}, V_{\delta P \delta P}\}}{2}.
\end{equation}

\section{Model of higher order mode}

In state preparation, higher-order mode noise detrimentally affects the conditional state~\cite{meng_measurement-based_2022}.  We model this effect by treating the higher-order mode as white noise. To understand this, we use the error spectrum of the non-causal Wiener filtering error $\Delta x = x - x_{\mathrm{est}}^{\mathrm{non}} $, which has the feature $\Delta x \sim \delta x$ and is given by~\cite{wiener_extrapolation_1964,bowen_quantum_2015,meng_optomechanical_2022}
\begin{equation}
	\label{eq:noncausal_error}
	S_{\Delta x \Delta x}(\omega)=\frac{S_{xx}(\omega)}{1+\mathrm{SNR}(\omega)},
\end{equation}
with the signal-to-noise ratio $\mathrm{SNR}(\omega)$ defined as $S_{xx}(\omega)/S_{nn}^{\mathrm{tot}}(\omega)$, where $S_{nn}^{\mathrm{tot}}=S_{xx}^{\mathrm{imp}}(\omega) + S_{xx,2}(\omega)$ and $S_{xx,2}(\omega)$ represents the PSD of the second-order mode noise. For high and low $\mathrm{SNR}(\omega)$ regimes, Eq.~(\ref{eq:noncausal_error}) simplifies to 
\begin{equation}
	S_{\Delta x \Delta x}(\omega) \approx 
	\begin{cases}
		&S_{nn}^{\mathrm{tot}}(\omega) \quad \text{if} \quad \mathrm{SNR}(\omega) \gg 1,
		\\
		&S_{xx}(\omega) \quad \text{if} \quad \mathrm{SNR}(\omega) \ll 1.
	\end{cases}
\end{equation}

\begin{figure}[h] \includegraphics[width=0.5\columnwidth]{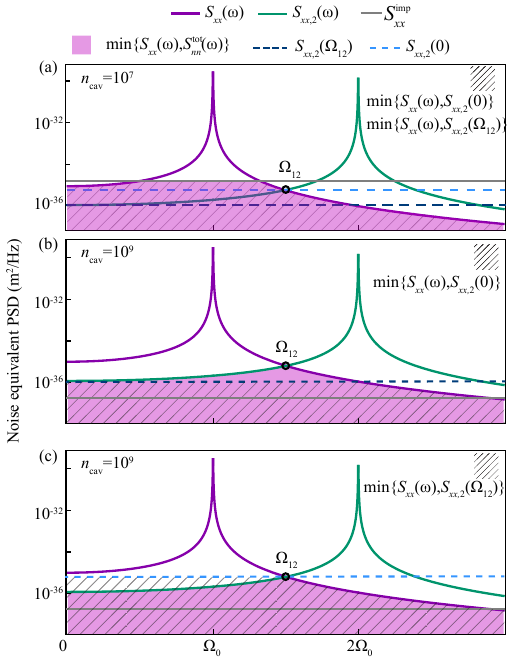}\caption{\label{fig:PSD}
	Noise equivalent power spectral density using the square membrane parameters. Purple line: signal from the fundamental mode.  Green line: noise from the second-order mode. Darkblue line: noise from the second-order mode at its zero frequency noise level. Lightblue line: noise from the second-order mode at its $\Omega_{12}$ frequency noise level. Grey line: optical shot noise. The optomechanical parameters are chosen as same as in the main text. (a) $S_{xx}(\Omega_{12})<S_{xx}^{\mathrm{imp}}$ for $n_{\mathrm{cav}}=10^{7}$. Pink shaded area: $\mathrm{min}\{S_{xx},S_{nn}^{\mathrm{tot}}\}$. Diagonal line pattern: the complete overlap of $\mathrm{min}\{S_xx,S_{xx}^{\mathrm{imp}}+S_{xx}(\Omega_{12})\}$ and $\mathrm{min}\{S_{xx},S_{xx}^{\mathrm{imp}}+S_{xx}(0)\}$. (b) and (c) $S_{xx}(\omega_{12})>S_{xx}^{\mathrm{imp}}$ for $n_{\mathrm{cav}}=10^{9}$. Diagonal line pattern: $\mathrm{min}\{S_{xx},S_{xx}^{\mathrm{imp}}+S_{xx}(0)\}$ for (b) and $\mathrm{min}\{S_{xx},S_{xx}^{\mathrm{imp}}+S_{xx}(\Omega_{12})\}$ for (c).
	}
\end{figure}

As a result, we can reexpress 
\begin{equation}
	S_{\Delta x \Delta x}(\omega) \approx \mathrm{min} \{ S_{xx}(\omega),S_{nn}^{\mathrm{tot}}(\omega)\}.
\end{equation}
The error spectrum is the minimum of either the signal from the fundamental mode or the total noise at a given frequency.  When the intracavity photon number is low that the condition $S_{xx,2}(\Omega_{12})<S_{xx}^{\mathrm{imp}}$ holds, where $\Omega_{12}$ represents the frequency at which the noise levels of the fundamental and second-order modes intersect, \textit{i.e.}, $\left.S_{xx}(\Omega_{12})=S_{xx,2}(\Omega_{12})\right.$. In this scenario, considering the noise from the second-order mode as white noise does not affect the error spectrum, as shown in Fig.~\ref{fig:PSD}~(a). However, as the intracavity photon number increases and reaches the regime where $S_{xx}(\Omega_{12})>S_{xx}^{\mathrm{imp}}$, treating the second-order mode noise as white at the magnitudes of $S_{xx,2}(0)$ and $S_{xx,2}(\Omega_{12})$ establishes the lower and upper bounds of the error spectrum, as illustrated in Fig.~\ref{fig:PSD}~(b) and (c), respectively.

\section*{Radiation pressure force and mechanical spring force}

In the main text, one of the feedback control mechanisms we consider is the radiation pressure force~\cite{cohadon_cooling_1999,arcizet_radiation-pressure_2006,kleckner_sub-kelvin_2006,meng_optomechanical_2022}. It is worth considering whether the radiation pressure force can be sufficiently strong to appreciably soften the oscillator. To evaluate this, given the measurement strength, we compare the radiation pressure force against the mechanical spring force. The latter arises from thermal and backaction fluctuations. These forces are quantified as $\left.F_{\text {rad }}=n_{\text {cav }} \hbar G\right.$ for the radiation pressure and $\left.F_{\text {mech }}=m \Omega_0^2 \sqrt{2 n_{\text {tot}}} x_{\mathrm{zp}}(\Omega_0)\right.$ for the mechanical spring force. The criterion for the radiation pressure force to exceed the mechanical spring force is established as $\left. \sqrt{2} n_{\text {cav }} g_0> \sqrt{n_{\text {tot}}}\Omega_0\right.$.

In the context of our study on the membrane-in-the-middle optomechanical device detailed in the main text, we find that  shifting the resonance frequency all the way to the zero requires only \( 9 \times 10^7 \) intracavity photons. This photon number is lower than $4\times 10^8$ used in the  experiment in Ref.~\cite{purdy_observation_2013}. This comparison underscores the feasibility of our approach in practical scenarios.


%\nocite{*}
%TC:ignore
\section*{References}
%\nocite{*}
%\footnotesize \renewcommand{\refname}{\vspace*{-30pt}}
\bibliographystyle{apsrev4-2} % Tell bibtex which bibliography style to use
\bibliography{Ref} % Tell bibtex which .bib file to use
%TC:endignore

\end{document}